\documentclass{article}

    \PassOptionsToPackage{numbers}{natbib}

    \usepackage[final]{neurips_2021}

\usepackage[utf8]{inputenc} 
\usepackage[T1]{fontenc}    
\usepackage{url}            
\usepackage{booktabs}       
\usepackage{amsfonts}       
\usepackage{nicefrac}       
\usepackage{microtype}      
\usepackage{xcolor}         
\usepackage{natbib}
\title{Artistic Autonomy in AI Art}

\author{
  Alayt Issak\\
  Coordinated Science Laboratory\\
  University of Illinois Urbana-Champaign\\
  Urbana, IL 61801 \\
  \texttt{aissak@illinois.edu} 
}

\begin{document}

\maketitle

\begin{abstract}
The concept of art has transposed meaning and medium across time, with its context being a deciding factor for its evolution. However, human beings' innermost functionality remains the same, and art, to this day, serves as an expression of the subconscious. Accelerated by the conception of GANs in 2014, automation has become a central medium in Artificial Intelligence (AI) Art. However, this raises concern over AI's influence on artistic autonomy within the process of creativity. This paper introduces the ethical responsibility of AI towards maintaining the artist’s volition in exercising autonomy and utilizes principles of self-determination theory alongside fundamental limits of creativity to do so. 
\end{abstract}

\section{Introduction: ethical care to creativity and intent}

The traditional role of automation in society served to make human lives easier by outsourcing mundane tasks. Recommender systems, for example, utilize language models to engage users in predictive text systems. However, much criticism has fallen on this medium as it alters the way people write. These systems have been found to make people “machine-like” – which is evident given its intention \cite{varshney_autonomy}. This prompts ethical care on the implementation of automation within attributes that characterize humanity – one of which is creativity. 

In psychoanalysis, creativity serves as the expressive element or natural human impulse that drives the artistic experience \cite{zweig_struggle}. It is what drives surprise within viewers for pushing the boundary of what is deemed to be the experience of reality. AI Art falls under criticism for automating this very process. For instance, as an agent of play to enact creativity, GANs are utilized as a black box for providing artistic result, where the feedback loop is based on the artist's alteration of the algorithm upon interpretation of results \cite{gans}. Unlike creation where artists decide meaning and form in process, AI Art limits artistic autonomy by basing the artist’s process upon output i.e. generating multiple sessions of training and determining the artwork based on generated artifacts (current exceptions go to CLIP with in-training modifications \cite{radford_clip}).

With regards to intent, GANs were originally focused on improving quality, stability, and variation \cite{radford_unsupervised} in order to implement the style transfer of the input image. Since then, they have evolved from representation to visually indeterminate artifacts to create an AI Art identity \cite{hertzmann_indeterminate}. However, the implementation of this medium still surrenders the creative process as the artifact's intent does not address the fundamental loss in autonomy that occurs within automation \cite{mccormack_monash}. In June 2021, a discussion series on AI research and social responsibility, titled Post-Human Creativity: The Use of AI in Art, featured artists who emphasized the need to strengthen "interactions between humans and machines... instead of making technology more human" as to preserve "meaningful interactions with algorithms and push the boundaries of creative processes." With the concerns for AI’s role in art in mind, we consider the ethical implications to the artist’s creative autonomy via principles in self-determination theory and intent via fundamental limits of creativity. 

\section{Defining creativity: self-determination theory}

Self-determination theory suggests that people are motivated to grow and change by three innate and universal psychological needs: autonomy, relatedness and competence \cite{ryan_self-determination}. Autonomy, or regulation by the self, is a phenomena that parallels other aspects of existence such as will, choice, and freedom. It is further augmented into liberty (independence from controlling principles) and agency (capacity for intentional action) \cite{define_autonomy}. 

We consider the limitation of AI Art to suffice liberty by considering emotion-based art. In the style transfer of AI Art, artists often use forms that acquire a sense of talent, such as impressionism, to replicate the delicacy of the form’s timeless novelty. However, in other forms such as Abstract expressionism, it is the human element of the artist that drives the form. Abstraction took time to develop appreciation due to the neglect for traditional talent \cite{abstract_resistance}, let alone expressionism, which is expressive of the artist's inner feelings \cite{expressionism}. This highlights the point in which creativity stems from the inner spark, or according to the psychoanalyst Carl Jung, "not accomplished by intellect but by play" or to a larger extent the "daimon of creativity" \cite{jung_symbols_1977}.

As such, Abstract expressionism is rooted in the creativity that spurs from the artist at the moment of creation. This moment, much like the deep immersion that comes with it, is encouraged and developed by a constant interaction that need not be interrupted, regulated, or automated \cite{diamond_anger_1996}. Hence, if one were to create AI Art based on Abstract expressionism, such as Jackson Pollock's action painting \cite{pollock}, then the result would lose its creative autonomy because of artistic interruption during the surrender of process to AI, as well as its core essence in conveying the emotion of the artist. 

\section{Defining intention: fundamental limits of creativity}

Intentionality is the inspiration or desire to express the human intent \cite{collingwood_intention}. The capacity for this action is captured by the need for agency in autonomy. Fundamental limits of creativity have detailed a limit theorem whereby tradeoff between novelty and quality for a given creative domain exists \cite{lav_math}. To consider a limit theorem for creativity with intentionality, Shannon’s capacity-cost-function formalism, which captures limits of reliable communication, is modified to address the semantic problem of creativity. Incorporating intention, semantic creativity shows that requiring communicative intent may reduce the quality and/or novelty of creative artifacts that are generated \cite{varshney_intentionality}. This inverse relationship between intent and novelty is paralleled by examples in Dada art, such as Duchamp's fountain, that, despite the utmost intent, garnered controversy on the novelty of artistic creation \cite{dada}. 

This begs to consider the role of novelty in AI Art due to the compromise in the intentional autonomy characterised by human creativity \cite{mccormack_monash}. One consideration would be to rethink novelty in AI Art and aim for a simultaneous increase of autonomy and intent. For instance, the DARCI (Digitial ARtist Communicating Intention) system builds an artificial system that exhibits creativity by noting the attribution of creativity with respect to system intentionality and autonomy. It has, thus far, addressed these difficulties and maintained to exhibit these characteristics to some extent \cite{ventura_autonomous_intention}. Drawing back to the “black box” analogy for AI training and the resultant novelty, one may consider the integration of intent within the system and assign the loss in novelty towards the artist. For example, the art collective aurèce vettier reinvents intent by exploring hybrid combinations of art and algorithms \cite{aurece2}. This way, with AI Art as a component, novelty arises out of the artist’s greater autonomy. 

\section{Conclusion}

The novelty of AI Art need not arise out of appreciation for AI's capability to create such works, but rather ask what the artwork entails in creativity and intent. 
It would be best to encourage artists to re-calibrate the role of AI in their art. One could explore their curiosity by incorporating room for play and perhaps unlock their inner creative in part of retreived autonomy within the process of creation.

\section{Acknowledgement}
Discussions with Lav R. Varshney are greatly appreciated.



\newpage

\end{document}